\documentclass[useAMS,usenatbib]{mn2e}
\usepackage[dvips]{graphicx}
\usepackage{longtable}

\title[ Photometric Observations of SN~2002er]{Photometric Observations of the Type Ia 
SN~2002er in UGC~10743}
\author[G. Pignata et al.]{G. Pignata,$^{1,2}$ F. Patat,$^1$ S. Benetti,$^3$ S. Blinnikov,$^{4,5}$ W. Hillebrandt,$^5$
\newauthor R. Kotak,$^6$ B. Leibundgut,$^1$ P. A. Mazzali,$^{7,5}$ P. Meikle,$^6$ Y. Qiu,$^8$ 
\newauthor  P. Ruiz-Lapuente,$^{5,9}$ S. Smartt,$^{10}$ E. Sorokina,$^{11}$ M. Stritzinger,$^5$
\newauthor   M. Stehle,$^{5,12}$   M. Turatto,$^3$  T. Marsh,$^{13}$  F. Martin-Luis,$^{14}$  N. McBride,$^{15}$ 
\newauthor J. Mendez,$^9$ L. Morales-Rueda,$^{13,16}$ D. Narbutis,$^{17}$ and R. Street$^{18}$   \\
$^1$ European Southern Observatory,  Karl-Schwarzschild-Str. 2, D-85748 Garching bei M\"unchen,  Germany.\\
$^2$ Dipartimento di Astronomia, Universit\'a di Padova, Vicolo dell'Osservatorio 2, I-35122 Padova, Italy \\
$^3$ Osservatorio Astronomico di Padova,Vicolo dell'Osservatorio 5, I-35122 Padova, Italy \\
$^4$ ITEP, 117218 Moscow, Russia \\
$^5$ Max-Planck-Institut  f\"ur Astrophysik, Karl-Schwarzschild-Str. 1, D-85741 Garching bei M\"unchen, Germany\\
$^6$ Blackett Laboratory, Imperial College London, Prince Consort Road, London SW7 2BW, UK\\
$^7$ Osservatorio Astronomico di Trieste, Via Tiepolo 11, I-34131 Trieste, Italy\\
$^8$ National Astronomical Observatories, Chinese Academy of Sciences, 100012 Beijing, China\\
$^9$ Department of Astronomy, University of Barcelona, Marti i Franques 1, E-08028 Barcelona, Spain\\
$^{10}$ Institute of Astronomy, University of Cambridge, Madingley Road, Cambridge CB3 0HA, UK\\
$^{11}$ Sternberg Astronomical Institute, Universitetski pr. 13, 119899  Moscow, Russia \\
$^{12}$ Universit\"ats-Sternwarte M\"unchen, Scheinerstr. 1, D-81679 M\"unchen, Germany\\
$^{13}$ Department of Physics and Astronomy, Southampton University, Southampton SO17 1BJ, UK\\
$^{14}$ Instituto de Astrof\'isica de Canarias, C/V\'ia L\'actea, s/n, 38205 La Laguna, Santa Cruz de Tenerife, Spain\\
$^{15}$ Planetary and Space Sciences Research Institute, The Open University, Milton Keynes, MK7 6AA, UK\\
$^{16}$ University of Nijmegen, 6500 GL Nijmegen, Netherlands\\
$^{17}$ Institute of Physics, Savanoriu pr. 231, LT-02300 Vilnius, Lithuania\\
$^{18}$ Department of Pure and Applied Physics, The Queen's University of Belfast, Belfast BT7 1NN, UK \\
 }

\begin{document}

\date{Accepted.......;  Received .......}

\pagerange{\pageref{firstpage}--\pageref{lastpage}} \pubyear{2002}

\maketitle

\label{firstpage}

\begin{abstract}
Extensive light and colour curves for the Type Ia supernova SN~2002er are
presented as part of the European Supernova Collaboration. We have collected
$UBVRI$ photometry from ten different telescopes covering the phases from 7 days
before until 619 days after maximum light. Corrections for the different
instrumental systems and the non-thermal spectrum of the supernova
(S-corrections) have been applied.

With the densely sampled light curves we can make detailed comparisons to
other well-observed objects. SN~2002er most closely resembles SN~1996X after
maximum, but clearly shows a different colour evolution before peak light and
a stronger shoulder in $V$ and $R$ bands compared to other well-observed SNe Ia. In
particular, the rise time appears to be longer than what is expected from
rise-time vs.decline-rate relation. 

We use several methods to determine the reddening towards SN~2002er based on
the colour evolution at near peak and at late phases. The $uvoir$ (bolometric)
light curve shows great similarity with SN~1996X, but also indications of a
higher luminosity, longer rise time and a more pronounced shoulder 25 days past maximum. 

The interpretation of the light curves was done with two independent light
curve codes. Both find that given the luminosity of SN~2002er the $^{56}$Ni
mass exceeds 0.6~M$_{\odot}$ with prefered values near 0.7~M$_{\odot}$. Uncertainties in the
exact distance to SN~2002er are the most serious limitation of this
measurement. The light curve modelling also indicates a high level of mixing of
the nickel in the explosion of SN~2002er.

\end{abstract}


\section{Introduction}

Even though significant progress has been made in recent years, many
of the properties of Type Ia Supernovae (SNe) remain largely
uncertain. To address this problem, several European institutes working in
this field have joined together to form the European Supernova
Collaboration (hereafter ESC, Benetti et al. 2004). One of the main
targets of this project is the collection of a large and homogeneous
database of optical and infrared SN Ia light curves and spectra at low redshift
($v_r~\leq$~6000~km~s$^{-1}$). A homogeneous and well-populated database is necessary in order to
search for systematic differences, evolution and/or environmental
effects, and for comparison with model predictions. All these
points are crucial in the use of SNe  Ia as cosmological probes.

 SN~2002er ( R.A.= $17^h 11^m 29^s.88$, DEC. = +$
7^\circ 59' 44''.8$, J2000) is one of the first targets of the ESC
project. It was discovered on 2002 August 23.2 UT in the spiral galaxy
UGC~10743 (Fig.~\ref{figure1}), $12''.3$ West and $4''.7$ North of
the galaxy nucleus, during the LOTOSS-KAIT SN search
\citep{Li}. On the basis of a low resolution spectrum taken at 
La Palma with the Isaac Newton Telescope (INT) on August 26.9 UT, the
candidate was classified by some of the ESC members as a type Ia SN,
approximately 10 days before maximum brightness \citep{Smartt}.

 Considering the early epoch of the discovery and the proximity of the
host galaxy ($v_r$~=~2568~$\pm$~7~km~s$^{-1}$, Falco et al. 1999),
target of opportunity observations were immediately triggered at all
available telescopes. Mainly thanks to the contribution of the Calar Alto
Observatory, almost daily coverage over the first month could be secured for both
photometry and spectroscopy.

In this paper we present the photometric results from 7 days before to
over 619 days past maximum light, including the analysis of the
bolometric light curve. Spectroscopic observations will be presented in a
separate paper (Kotak et al., in preparation).

\begin{table*}
\begin{minipage}{170mm}
\caption{Original photometric observations of SN~2002er.} \label{somemode}
\begin{tabular}{@{}cccccccccc}
\hline
date & M.J.D. & Phase$^a$ & U & B & V & R & I & Instr. & Seeing\\
& & (days) & & & & & & & (arcsec)\\
\hline
23/08/02 & 52509.2 & $-$15.0 &  $-$    &  $-$    &  $-$    &   17.5$^b$   &  $-$     & KAIT & $-$\\
24/08/02 & 52510.2 & $-$14.0 &  $-$    &  $-$    &  $-$    &   16.9$^b$   &  $-$     & KAIT & $-$\\
30/08/02 & 52516.8 &  $-$7.3 & $-$ & 15.48 $\pm$ 0.04 & 15.26 $\pm$ 0.04 & 15.00 $\pm$ 0.04 & 14.95 $\pm$ 0.04 & AF & 1.8 \\
30/08/02 & 52516.8 &  $-$7.3 & $-$ & $-$ & $-$ & 14.96 $\pm$ 0.04 & 14.94 $\pm$ 0.04 & WM & 1.5 \\
30/08/02 & 52516.9 &  $-$7.2 & 15.15 $\pm$ 0.09 & 15.48 $\pm$ 0.04 & 15.25 $\pm$ 0.03 & 14.96 $\pm$ 0.04 & 14.91 $\pm$ 0.04 & CS & 1.6 \\
31/08/02 & 52517.9 &  $-$6.2 & 15.02 $\pm$ 0.10 & 15.30 $\pm$ 0.04 & 15.11 $\pm$ 0.03 & 14.82 $\pm$ 0.03 & 14.76 $\pm$ 0.04 & CS & 1.7 \\
31/08/02 & 52518.0 &  $-$6.1 & $-$ & 15.36 $\pm$ 0.05 & 15.08 $\pm$ 0.04 & 14.80 $\pm$ 0.04 & 14.74 $\pm$ 0.05 & JJ & 2.7 \\
01/09/02 & 52518.9 &  $-$5.2 & 14.98 $\pm$ 0.05 & 15.15 $\pm$ 0.04 & 14.96 $\pm$ 0.03 & 14.66 $\pm$ 0.03 & 14.63 $\pm$ 0.04 & CS & 1.9 \\
02/09/02 & 52519.9 &  $-$4.2 & 14.83 $\pm$ 0.03 & 15.10 $\pm$ 0.03 & 14.85 $\pm$ 0.03 & 14.58 $\pm$ 0.04 & $-$ & CL & 1.7 \\
03/09/02 & 52520.9 &  $-$3.2 & 14.82 $\pm$ 0.03 & 14.99 $\pm$ 0.03 & 14.83 $\pm$ 0.03 & 14.56 $\pm$ 0.03 & 14.51 $\pm$ 0.03 & CL & 3.5 \\
04/09/02 & 52521.9 &  $-$2.3 & 14.73 $\pm$ 0.03 & 14.92 $\pm$ 0.03 & 14.75 $\pm$ 0.03 & 14.49 $\pm$ 0.03 & 14.48 $\pm$ 0.03 & CL & 2.1 \\
05/09/02 & 52522.8 &  $-$1.4 & $-$ & $-$ & $-$ & 14.45 $\pm$ 0.03 & 14.48 $\pm$ 0.03 & WM & 1.8 \\
05/09/02 & 52522.9 &  $-$1.3 & 14.74 $\pm$ 0.04 & 14.90 $\pm$ 0.03 & 14.70 $\pm$ 0.03 & 14.47 $\pm$ 0.03 & 14.45 $\pm$ 0.03 & CL & 2.5 \\
06/09/02 & 52523.8 &  $-$0.4 & $-$ & $-$ & $-$ & 14.46 $\pm$ 0.03 & $-$ & WM & 2.6 \\
06/09/02 & 52523.9 &  $-$0.3 & 14.72 $\pm$ 0.03 & 14.87 $\pm$ 0.03 & 14.65 $\pm$ 0.03 & 14.45 $\pm$ 0.03 & 14.47 $\pm$ 0.03 & CL & 1.3 \\
07/09/02 & 52524.9 &   ~~0.8 & 14.74 $\pm$ 0.03 & 14.91 $\pm$ 0.03 & 14.63 $\pm$ 0.03 & 14.43 $\pm$ 0.03 & 14.46 $\pm$ 0.03 & CL & 3.3 \\
08/09/02 & 52525.9 &   ~~1.8 & 14.79 $\pm$ 0.03 & 14.91 $\pm$ 0.03 & 14.64 $\pm$ 0.03 & 14.44 $\pm$ 0.03 & 14.50 $\pm$ 0.03 & CL & 3.1 \\
10/09/02 & 52527.9 &   ~~3.7 & 14.86 $\pm$ 0.08 & 14.99 $\pm$ 0.04 & 14.62 $\pm$ 0.04 & 14.46 $\pm$ 0.04 & 14.59 $\pm$ 0.04 & CS & 1.6 \\
12/09/02 & 52530.0 &   ~~5.8 & 15.09 $\pm$ 0.07 & $-$ & 14.68 $\pm$ 0.04 & 14.55 $\pm$ 0.03 & 14.66 $\pm$ 0.05 & CS & 2.2 \\
13/09/02 & 52530.9 &   ~~6.7 & 15.29 $\pm$ 0.04 & 15.21 $\pm$ 0.03 & 14.69 $\pm$ 0.03 & 14.62 $\pm$ 0.04 & 14.82 $\pm$ 0.04 & JJ & 1.2 \\
14/09/02 & 52531.5 &   ~~7.4 & $-$ & $-$ & 14.72 $\pm$ 0.03 & 14.68 $\pm$ 0.04 & 14.88 $\pm$ 0.04 & BA & 2.9 \\
14/09/02 & 52531.9 &   ~~7.8 & 15.40 $\pm$ 0.07 & 15.36 $\pm$ 0.03 & 14.78 $\pm$ 0.03 & 14.72 $\pm$ 0.04 & 14.92 $\pm$ 0.04 & JJ & 1.7 \\
16/09/02 & 52533.5 &   ~~9.3 & $-$ & 15.54 $\pm$ 0.07 & 14.85 $\pm$ 0.03 & 14.80 $\pm$ 0.04 & 14.99 $\pm$ 0.03 & BA & 2.4 \\
16/09/02 & 52533.9 &   ~~9.8 & $-$ & 15.56 $\pm$ 0.04 & 14.91 $\pm$ 0.04 & 14.92 $\pm$ 0.04 & 15.08 $\pm$ 0.04 & JJ & 1.8 \\
18/09/02 & 52535.0 &  ~~10.9 & 16.00 $\pm$ 0.10 & 15.64 $\pm$ 0.03 & 14.98 $\pm$ 0.03 & 14.94 $\pm$ 0.04 & 15.22 $\pm$ 0.04 & DF & 1.5 \\
20/09/02 & 52537.9 &  ~~13.7 & $-$ & 15.98 $\pm$ 0.04 & $-$ & $-$ & $-$ & JJ & 2.4 \\
21/09/02 & 52538.5 &  ~~14.3 & $-$ & 16.12 $\pm$ 0.04 & 15.15 $\pm$ 0.03 & 15.11 $\pm$ 0.04 & 15.10 $\pm$ 0.03 & BA & 2.3 \\
21/09/02 & 52538.8 &  ~~14.7 & $-$ & $-$ & $-$ & $-$ & 15.11 $\pm$ 0.04 & JJ & 1.7 \\
22/09/02 & 52539.9 &  ~~15.7 & $-$ & 16.27 $\pm$ 0.04 & $-$ & $-$ & 15.12 $\pm$ 0.04 & JJ & 1.0 \\
22/09/02 & 52540.0 &  ~~15.8 & 16.76 $\pm$ 0.05 & 16.28 $\pm$ 0.03 & 15.29 $\pm$ 0.03 & 15.13 $\pm$ 0.03 & 15.21 $\pm$ 0.03 & DF & 2.0 \\
23/09/02 & 52540.9 &  ~~16.7 & $-$ & 16.39 $\pm$ 0.04 & $-$ & $-$ & 15.07 $\pm$ 0.04 & JJ & 0.9 \\
24/09/02 & 52542.0 &  ~~17.8 & 16.97 $\pm$ 0.08 & 16.51 $\pm$ 0.04 & 15.42 $\pm$ 0.04 & $-$ & $-$ & DF & 1.0 \\
26/09/02 & 52544.0 &  ~~19.9 & 17.24 $\pm$ 0.08 & 16.74 $\pm$ 0.03 & 15.53 $\pm$ 0.03 & 15.17 $\pm$ 0.04 & 15.12 $\pm$ 0.04 & DF & 1.1 \\
27/09/02 & 52544.9 &  ~~20.7 & $-$ & $-$ & 15.55 $\pm$ 0.03 & $-$ & $-$ & JJ & 1.1 \\
28/09/02 & 52545.9 &  ~~21.7 & $-$ & $-$ & 15.60 $\pm$ 0.03 & $-$ & $-$ & JJ & 1.2 \\
29/09/02 & 52546.5 &  ~~22.3 & $-$ & 17.04 $\pm$ 0.05 & 15.58 $\pm$ 0.03 & 15.21 $\pm$ 0.04 & 14.95 $\pm$ 0.03 & BA & 2.0 \\
29/09/02 & 52546.8 &  ~~22.6 & $-$ & $-$ & $-$ & 15.27 $\pm$ 0.04 & 15.01 $\pm$ 0.04 & WM & 1.5 \\
29/09/02 & 52546.9 &  ~~22.7 & $-$ & $-$ & 15.64 $\pm$ 0.03 & $-$ & $-$ & JJ & 0.9 \\
30/09/02 & 52547.8 &  ~~23.6 & $-$ & $-$ & $-$ & 15.26 $\pm$ 0.03 & 14.98 $\pm$ 0.04 & WM & 1.6 \\
30/09/02 & 52547.8 &  ~~23.6 & $-$ & 17.10 $\pm$ 0.03 & 15.71 $\pm$ 0.03 & 15.26 $\pm$ 0.04 & 14.98 $\pm$ 0.03 & AF & 1.7 \\
01/10/02 & 52548.9 &  ~~24.7 & $-$ & $-$ & 15.72 $\pm$ 0.03 & 15.32 $\pm$ 0.03 & 14.93 $\pm$ 0.03 & JJ & 1.8 \\
04/10/02 & 52551.5 &  ~~27.3 & $-$ & 17.45 $\pm$ 0.07 & 15.89 $\pm$ 0.04 & 15.40 $\pm$ 0.04 & 14.95 $\pm$ 0.04 & BA & 2.8 \\
05/10/02 & 52552.5 &  ~~28.3 & $-$ & 17.59 $\pm$ 0.06 & 15.99 $\pm$ 0.04 & 15.46 $\pm$ 0.04 & 15.04 $\pm$ 0.04 & BA & 3.2 \\
07/10/02 & 52554.8 &  ~~30.7 & $-$ & 17.64 $\pm$ 0.03 & 16.19 $\pm$ 0.04 & 15.70 $\pm$ 0.05 & 15.17 $\pm$ 0.05 & MO & 0.9 \\
08/10/02 & 52555.8 &  ~~31.6 & $-$ & $-$ & $-$ & 15.77 $\pm$ 0.04 & 15.29 $\pm$ 0.04 & WM & 2.7 \\
08/10/02 & 52555.8 &  ~~31.6 & 17.85 $\pm$ 0.03 & 17.71 $\pm$ 0.04 & 16.27 $\pm$ 0.04 & 15.75 $\pm$ 0.03 & 15.25 $\pm$ 0.03 & TD & 1.2 \\
09/10/02 & 52556.8 &  ~~32.6 & $-$ & $-$ & $-$ & 15.82 $\pm$ 0.04 & 15.43 $\pm$ 0.05 & WM & 3.0 \\
13/10/02 & 52560.4 &  ~~36.2 & $-$ & $-$ & 16.51 $\pm$ 0.04 & 16.04 $\pm$ 0.04 & 15.71 $\pm$ 0.04 & SS & 2.7 \\
14/10/02 & 52561.4 &  ~~37.2 & $-$ & 17.93 $\pm$ 0.05 & 16.58 $\pm$ 0.04 & $-$ & $-$ & SS & 2.0 \\
16/10/02 & 52563.8 &  ~~39.6 & $-$ & $-$ & $-$ & 16.23 $\pm$ 0.04 & 15.89 $\pm$ 0.06 & WM & 2.4 \\
26/10/02 & 52573.4 &  ~~49.3 & $-$ & 18.20 $\pm$ 0.15 & 16.95 $\pm$ 0.05 & 16.66 $\pm$ 0.08 & 16.29 $\pm$ 0.14 & BA & 3.7 \\
27/10/02 & 52574.4 &  ~~50.3 & $-$ & $-$ & 17.01 $\pm$ 0.06 & $-$ & 16.35 $\pm$ 0.08 & BA & 4.0 \\
04/03/03 & 52703.4 & ~~179.2 & $-$ & 19.94 $\pm$ 0.03 & 19.90 $\pm$ 0.05 & 20.10 $\pm$ 0.09 & 19.78 $\pm$ 0.16 & WF & 1.5 \\
28/03/03 & 52726.3 & ~~202.2 & $-$ & 20.28 $\pm$ 0.05 & 20.17 $\pm$ 0.04 & 20.55 $\pm$ 0.05 & 20.20 $\pm$ 0.07 & WF & 1.1 \\
09/04/03 & 52739.3 & ~~215.2 & $-$ & 20.48 $\pm$ 0.06 & 20.37 $\pm$ 0.05 & $-$ & $-$ & WF & 1.6 \\
10/04/03 & 52740.3 & ~~216.2 & $-$ & $-$ & $-$ & 20.85 $\pm$ 0.07 & 20.38 $\pm$ 0.06 & WF & 0.8 \\
29/07/03 & 52848.9 & ~~324.7 & $-$ & 21.84 $\pm$ 0.06 & $-$ & $-$ & $-$ & MO & 1.1 \\
18/05/04 & 53143.3 & ~~619.1 & $-$ & $>$ 23.1 & $>$ 23.7  & $-$ & $-$ & VF & 0.9 \\
\hline
\multicolumn{10}{l}{($^a$) Relative to $B$ maximum (MJD= 52524.2)}\\
\multicolumn{10}{l}{($^b$) Unfiltered magnitude reasonably well approximated to R band magnitude (Li, private communication)}\\
\multicolumn{10}{l}{CS = Calar Alto 2.2m + CAFOS + CCD SITe 0\farcs53$/pix$; DF = Danish 1.54m + DFOSC 0\farcs39$/pix$; BA = BAO 0.85m + CCD 0\farcs45$/pix$}\\
\multicolumn{10}{l}{ CL = Calar Alto 2.2m + CAFOS + CCD Loral 0\farcs33$/pix$; WF = ESO 2.2m + WFI 0\farcs24$/pix$; VF = ESO VLT + FORS1 0\farcs20$/pix$ }\\
\multicolumn{10}{l}{JJ = JKT 1.0m + JAG 0\farcs33$/pix$; WM = Wendelstein 0.8m + MONICA 0\farcs50$/pix$; AF = Asiago 1.82m + AFOSC 0\farcs34$/pix$}\\
\multicolumn{10}{l}{SS = Siding Spring 2.3.m + MSSSO11 0\farcs59$/pix$; TD = TNG + DOLORES 0\farcs28$/pix$; MO = Calar Alto 3.5m + MOSCA 0\farcs32$/pix$} \\
\end{tabular}
\end{minipage}
\label{table1}
\end{table*}

\section{Observations and Data Reduction}

\begin{table*}
\begin{minipage}{110mm}
\caption{Magnitudes for the local sequence stars in the field of SN~2002er (Fig. \ref{figure1}). The data were obtained on 10 photometric nights with CAFOS, DFOSC and  WFI.}
\begin{tabular}{@{}cccccc}
  \hline id & U & B & V & R & I \\
\hline
01 & 17.97 $\pm$ 0.03  & 17.14 $\pm$ 0.03 & 16.06 $\pm$ 0.03 & 15.45 $\pm$ 0.03 & 14.90 $\pm$ 0.02 \\ 
02 &        $-$        & 18.12 $\pm$ 0.04 & 17.10 $\pm$ 0.03 & 16.47 $\pm$ 0.02 & 15.92 $\pm$ 0.02 \\ 
03 & 15.93 $\pm$ 0.02  & 15.55 $\pm$ 0.01 & 14.70 $\pm$ 0.02 & 14.20 $\pm$ 0.03 & 13.75 $\pm$ 0.03 \\
04 & 17.03 $\pm$ 0.02  & 16.75 $\pm$ 0.02 & 15.92 $\pm$ 0.03 & 15.44 $\pm$ 0.03 & 14.99 $\pm$ 0.02 \\
05 & 17.12 $\pm$ 0.03  & 16.58 $\pm$ 0.02 & 15.55 $\pm$ 0.01 & 14.98 $\pm$ 0.03 & 14.42 $\pm$ 0.03 \\
06 & 18.01 $\pm$ 0.02  & 16.84 $\pm$ 0.02 & 15.30 $\pm$ 0.02 & 14.23 $\pm$ 0.05 & 13.03 $\pm$ 0.04 \\
07 &        $-$        & 19.05 $\pm$ 0.02 & 18.32 $\pm$ 0.04 & 17.85 $\pm$ 0.03 & 17.42 $\pm$ 0.04 \\
08 &        $-$        & 19.31 $\pm$ 0.04 & 18.44 $\pm$ 0.04 & 17.94 $\pm$ 0.03 & 17.48 $\pm$ 0.05 \\
09 &        $-$        & 18.38 $\pm$ 0.03 & 17.31 $\pm$ 0.04 & 16.68 $\pm$ 0.03 & 16.13 $\pm$ 0.05 \\
10 & 18.54 $\pm$ 0.06  & 18.24 $\pm$ 0.04 & 17.39 $\pm$ 0.04 & 16.90 $\pm$ 0.04 & 16.48 $\pm$ 0.02 \\
11 & 18.64 $\pm$ 0.05  & 18.62 $\pm$ 0.02 & 17.92 $\pm$ 0.05 & 17.47 $\pm$ 0.04 & 17.01 $\pm$ 0.07 \\
12 &        $-$        & 17.87 $\pm$ 0.03 & 16.66 $\pm$ 0.03 & 15.91 $\pm$ 0.02 & 15.29 $\pm$ 0.04 \\
\hline
\end{tabular}
\end{minipage}
\label{table2}
\end{table*}

\begin{figure}
\includegraphics[width=84mm]{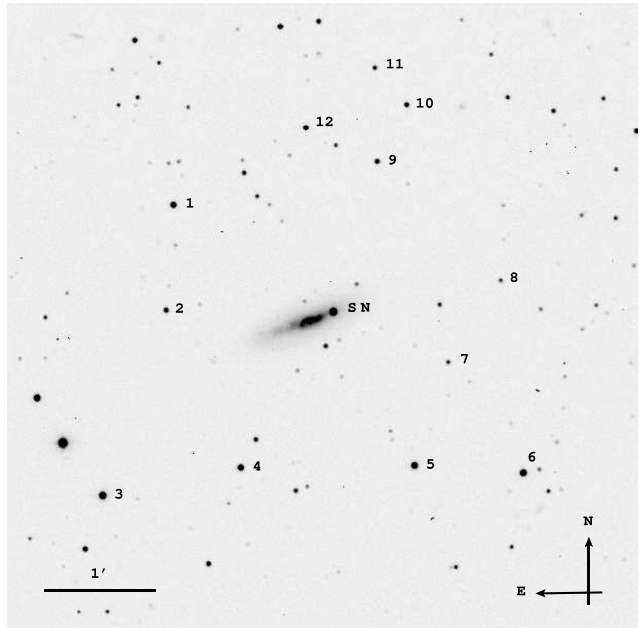}
\caption{Identification chart for SN~2002er and the local sequence stars 
(V-band exposure obtained on 2003, September 6 with the 2.2m+CAFOS).}
\label{figure1}
\end{figure}

The observations of SN~2002er were performed with twelve different
facilities. The main parameters of each instrument are summarized in
the notes of Table~\ref{table1}.  Basic data reduction (bias and
flat-field correction) was performed using standard routines in
IRAF\footnote{IRAF is distributed by the National Optical Astronomy
Observatories, which are operated by the Association of Universities
for Research in Astronomy, Inc, under contract to the National Science
Foundation.}.  Selected observations of standard fields
\citep{Landolt} during photometric nights have been used to compute
the colour terms for each instrument and to calibrate a local standard
sequence, which in turn, was used to calibrate frames obtained under
non-photometric conditions.  The stars of the local sequence are
identified in Fig.~\ref{figure1} and their magnitudes are listed in
Table~\ref{table2} along with estimated root mean square (RMS) errors, which are typically
on the order of 0.03 mag.  Since the SN was projected
onto a region with quite a complicated background, photometry was
performed using the PSF fitting technique.  Uncertainties for the
magnitude measurements were estimated by combining the
fit errors in quadrature with those introduced by the transformation of instrumental
magnitudes into the $UBVRI$ Johnson-Cousin standard system
\citep{Bessell}.

\section{Photometric systems characterisation}

\begin{table*}
\begin{minipage}{110mm}
 \caption{Comparison between synthetic and photometric colour terms.}
 \label{symbols}
 \begin{tabular}{@{}lcccccccc}
  \hline
\multicolumn{9}{l}{Facilities ~~~~~~~~~~~~~~~  $B$~$B-V$ ~~~~~~~~~~~~~  $V$~$B-V$  ~~~~~~~~~~~~~ $R$~$R-I$  ~~~~~~~~~~~~~  $I$~$R-I$}\\
  & sy & ph & sy & ph & sy & ph & sy & ph \\
\hline
CAFOS+Loral & ~~0.129   &  ~~0.108  & $-$0.073 & $-$0.069 & $-$0.011 & $-$0.062 &  ~~0.209 &  ~~0.236 \\
CAFOS+Site  & ~~0.122   &  ~~0.120  & $-$0.067 & $-$0.052 &  ~~0.015 & $-$0.021 &  ~~0.231 &  ~~0.209 \\
DOLORES     & ~~0.064   &  ~~0.077  & $-$0.115 & $-$0.108 &  ~~0.041 &  ~~0.026 &  ~~0.020 &  ~~0.013 \\
DFOSC 	    & ~~0.063   &  ~~0.089  &  ~~0.008 &  ~~0.010 &  ~~0.016 &  ~~0.013 & $-$0.040 & $-$0.044 \\
JKT         & ~~0.043   &  ~~0.055  &  ~~0.032 &  ~~0.038 &  ~~0.014 & $-$0.009 &  ~~0.060 &  ~~0.047 \\
AFOSC       &$-$0.015 & $  -$0.040  &  ~~0.054 &  ~~0.060 &  ~~0.074 &  ~~0.060 & $-$0.042 & $-$0.040 \\
MOSCA       & ~~0.212   &  ~~0.219  &  ~~0.034 &  ~~0.030 & $-$0.007 & $-$0.018 &  ~~0.164 &  ~~0.140 \\
SSO         & ~~0.030   &  ~~0.022  &  ~~0.008 &  ~~0.015 &  ~~0.031 &  ~~0.015 & $-$0.030 & $-$0.035 \\
WFI         & ~~0.226   &  ~~0.242  & $-$0.070 & $-$0.068 &  ~~0.024 &  ~~0.010 &  ~~0.018 &  ~~0.038 \\
\hline
 \end{tabular}
\end{minipage}
\label{table3}
\end{table*}

It is well known that different instrumental photometric systems never
 exactly match. It is also known that the colour equations
used to transpose the instrumental magnitude into a standard system
loose their accuracy for objects characterised by a non-stellar
spectrum.  Since the photometry of SN~2002er was obtained using almost
all the instruments available to the ESC and these facilities will be
used to monitor other targets in the future, we have decided to
perform a detailed characterisation of the photometric properties of each
instrument. This in fact allows us to correct the systematic photometric errors introduced by the combination of the caveats just  mentioned before, using the S-correction method presented by \citet{Stritzinger} and \citet{Krisciunas1}.
In order to compute the corrections, one needs first to determine the
instrumental passband $S(\lambda)$:\\

$S(\lambda)=F(\lambda)\cdot QE(\lambda) \cdot A(\lambda) \cdot M(\lambda) \cdot L(\lambda)$\\

\noindent where $F(\lambda)$ is the filter transmission function, $QE(\lambda)$ is the detector quantum efficiency, $A(\lambda)$ is the continuum atmospheric transmission profile, $M(\lambda)$ is the mirror reflectivity function and  $L(\lambda)$ is the
lens throughput.  We determined $S(\lambda)$ for all the facilities for
which we could obtain the previously mentioned information. i.e. CAFOS+Loral,
CAFOS+SITe, JKT, AFOSC, DFOSC, MOSCA, DOLORES, 2.3m SSO and WFI.
$F(\lambda)$ and  $QE(\lambda)$ were usually
downloaded from the instrument web sites. We obtain $A(\lambda)$  for Calar Alto and La Palma in
\citet{atm_caha} and \citet{atm_lapalma} respectively, while for La Silla we
have used the CTIO transmission curve in the IRAF reduction package. Finally, for
Asiago and Siding Spring Observatory we obtained $A(\lambda)$ by modifying the standard atmospheric model
proposed by \citet{Walker} in order to match the average broad band
absorption coefficients of those sites. For $M(\lambda)$ we have used a standard aluminium reflectivity curve.
For the lenses several materials are generally used, but the
corresponding  $L(\lambda)$ are relatively flat across the optical range. Since it is very difficult, and often not
possible to get this information for all the instruments used,
we have decided to assume that  $L(\lambda)$ is constant
across the whole spectral range. This might cause some problems
at wavelengths bluer than 3400~\AA, i.e. affecting the $U$ passband.
However we have decided  not to determine $S(\lambda)$ for the  U filter, since
very few SN spectra cover this wavelength range,  making it very difficult to compute the relative S-correction.  

Once the response functions are constructed, one needs to calculate an
instrumental zero point for each passband. For this purpose we have
used a subset of spectrophotometric standard stars (Hamuy
\citealt{Hamuy_1},\citealt{Hamuy_2}), for which photoelectric
photometry is also available
(Cousins\citealt{Cousins_a},\citealt{Cousins_b},\citealt{Cousins_c}).
In order to check how well the modelled passbands match the real ones
we derived synthetic colour terms using the full set of Hamuy
spectrophotometric standards. As the reader can see in
Table~\ref{table3}, the match with the colour terms computed using the
classical photometric method is quite good for all instruments,
especially in the $V$ band. Of course, this is not a definitive proof
that the reconstructed passbands perfectly resemble the real ones,
since the stellar spectra, which are dominated by a smooth continuum,
are not the best tool to highlight differences between  photometric
systems. However, the tight correspondence shown by this test
makes us confident that the reproduction is reasonably accurate.

\begin{figure}
\includegraphics[width=84mm]{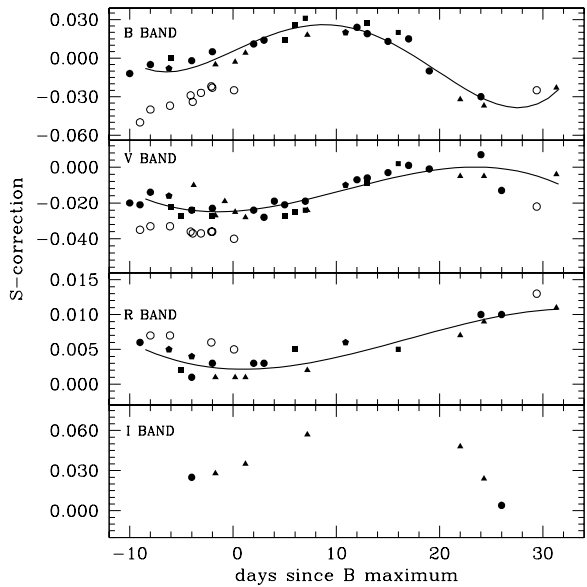}
\caption{Evolution of the S-correction in the $B$, $V$, $R$, 
and $I$ bands for 2.2m+CAFOS which was the most used facility during the
observational campaign. The solid line represents the fitted
polynomial, while the different symbols refer to SN~1992A (filled squares), SN~1994D (filled circles), SN~1996X (filled triangles), SN~2002er (filled pentagons) and SN~2002bo (open circles).}
\label{figure2}
\end{figure}

With this information we can  check the accuracy of our photometry and use the S-correction technique in order to remove possible systematic errors.  With the aim of computing a
reliable correction for the whole phase range covered by the
photometry, we decided to use not only the best flux calibrated
spectra of SN~2002er, but also a set of spectra from SN~1992A
(ESO-Key Project, unpublished), SN~1994D \citep{Patat} and SN~1996X \citep{Salvo},
properly reddened in order to match the continuum of SN~2002er.
The spectra were also shifted to the  SN~2002er rest frame. These
SNe have very similar spectra at every phase and as the reader can
see in Fig.~\ref{figure2}, this makes the S-correction reasonably
similar. Moreover, the spectra of different SNe are uniformly
distributed along the light curve, preventing possible systematic
errors.
The similarity between the used spectra is a crucial point. To prove it, in Fig. 2 we have also plotted the corrections computed for another ESC target, SN~2002bo \citep{Benetti}, that is a SN characterised by  different features and reddening. As the reader can easily note, especially in the B band, the corrections are significantly different from those of the other SNe. To describe the behaviour of the S-correction as a function
of time, for each instrument we have fitted a third or fourth order
polynomial to the data points. The RMS deviation of the observed
points from the fitted law gives us an estimate of the error
associated with the correction itself. This includes both the
uncertainties relative to the flux calibration of the input spectra
and the noise generated by using data from different SNe. For the $I$
band, where the phase coverage is quite scanty, the correction was
obtained by linear interpolation. For this filter the uncertainties
are larger than in $B$,$V$ and $R$. This is due both to the sparse
sampling and to the fact that the sensitivity of most spectrographs
falls toward the red edge of the $I$ band, making the flux
calibration, and in turn, the S-correction less reliable.

A further way to check the effectiveness of the S-correction is the
comparison between the data spread before and after its application.
The RMS deviation from a fitted polynomial in $B$ and $V$ is reduced
by a 0.001 and 0.002 magnitudes respectively, while for the $R$ filter
it does not change.  This negligible improvement simply means that the
original photometry was substantially accurate. The correction in
those bands is in fact quite small (see Table~\ref{table4}), and the
observed scatter is dominated by the random photometric error.
Nevertheless, our main goal was to check and eventually remove
systematic effects introduced by the instrumental photometric system.
For example, the peak brightness of SN~2002er was covered mainly with
CAFOS and during this phase range the S-correction for CAFOS $B$ and $V$
bands was +0.01 and $-$0.025 respectively. This makes the SN
0.035 mag bluer than it really is, and this has some impact, for
instance, on the reddening estimate.

Generally, the $I$ passband departs from the standard one more than in all other filters, and in
turn, the corrections are larger. In this case, the systematic errors
introduced by the natural photometric system mismatch are comparable to,
or larger than, random errors. Therefore, after the correction, the
RMS falls to a more meaningful value (0.017 mag).  We note that
this mismatch between the natural systems could be one of the reasons
why the $I$ light curves of type Ia SNe usually show differences which
are more pronounced than in any other filter (see for example Suntzeff
1996).\\ For the 0.85m Beijing and 0.80m Wendelstein
telescopes it was not possible to find the basic information needed to
compute the global response. Since the original photometry obtained at
these two observatories was not too different from that of the others,
we have conservatively applied the closest in time correction derived
for the observations carried out with known passbands.  The magnitudes
reported in Table~\ref{table1} are the original measurements but all
the relevant parameters like $E(B-V)$, $\Delta m_{15}$ and so on, were
computed using the corrected data.

\begin{table*}
\begin{minipage}{110mm}
 \caption{S-corrections to be added to the data in Table~\ref{table1}
 to bring them in the photometric standard system \citet{Bessell}. The
 meaning of the acronyms in the last column is explained in the notes
 of Table~\ref{table1}.}
\label{table4}
 \begin{tabular}{@{}lcccccc}
  \hline
MJD  &  Phase$^a$  & $B$ & $V$  & $R$  & $I$  & Facilities  \\
& (days) &  & & & & \\
\hline
52516.8 &   $-$7.3 & $-$0.010 $\pm$ 0.006 &  0.004 $\pm$ 0.005 & $-$0.003 $\pm$ 0.002 &  ~~0.036 $\pm$ 0.012& AF\\
52516.8 &   $-$7.3 & $-$ & $-$ &  ~~0.005 & ~~0.016& WM\\
52516.9 &   $-$7.2 & $-$0.012 $\pm$ 0.007 & $-$0.021 $\pm$ 0.005 &  ~~0.004 $\pm$ 0.002 &  ~~0.016 $\pm$ 0.014& CS\\
52517.9 &   $-$6.2 & $-$0.012 $\pm$ 0.007 & $-$0.023 $\pm$ 0.005 &  ~~0.004 $\pm$ 0.002 &  ~~0.019 $\pm$ 0.014& CS\\
52518.0 &   $-$6.1 & $-$0.020 $\pm$ 0.006 &  ~~0.005 $\pm$ 0.006 &  ~~0.002 $\pm$ 0.001 &  ~~0.042 $\pm$ 0.012& JJ\\
52518.9 &   $-$5.2 & $-$0.011 $\pm$ 0.007 & $-$0.024 $\pm$ 0.005 &  ~~0.003 $\pm$ 0.002 &  ~~0.021 $\pm$ 0.014& CS\\
52519.9 &   $-$4.2 & $-$0.009 $\pm$ 0.007 & $-$0.025 $\pm$ 0.005 &  ~~0.003 $\pm$ 0.002 & $-$& CL\\
52520.9 &   $-$3.2 & $-$0.007 $\pm$ 0.007 & $-$0.025 $\pm$ 0.005 &  ~~0.003 $\pm$ 0.002 &  ~~0.026 $\pm$ 0.014& CL\\
52521.9 &   $-$2.3 & $-$0.004 $\pm$ 0.007 & $-$0.026 $\pm$ 0.005 &  ~~0.002 $\pm$ 0.002 &  ~~0.028 $\pm$ 0.014& CL\\
52522.8 &   $-$1.4 & $-$ & $-$ &  ~~0.002 &  ~~0.030 & WM\\
52522.9 &   $-$1.3 & $-$0.000 $\pm$ 0.007 & $-$0.026 $\pm$ 0.005 &  ~~0.002 $\pm$ 0.002 &  ~~0.030 $\pm$ 0.014& CL\\
52523.8 &   $-$0.4 & $-$ & $-$ &  ~~0.002 & $-$& WM\\
52523.9 &   $-$0.3 &  ~~0.004 $\pm$ 0.007 & $-$0.026 $\pm$ 0.005 &  ~~0.002 $\pm$ 0.002 &  ~~0.032 $\pm$ 0.014& CL\\
52524.9 &   ~~0.8 &  ~~0.008 $\pm$ 0.007 & $-$0.025 $\pm$ 0.005 &  ~~0.002 $\pm$ 0.002 &  ~~0.035 $\pm$ 0.014& CL\\
52525.9 &   ~~1.8 &  ~~0.012 $\pm$ 0.007 & $-$0.024 $\pm$ 0.005 &  ~~0.002 $\pm$ 0.002 &  ~~0.037 $\pm$ 0.014& CL\\
52527.9 &   ~~3.7 &  ~~0.019 $\pm$ 0.007 & $-$0.023 $\pm$ 0.005 &  ~~0.002 $\pm$ 0.002 &  ~~0.041 $\pm$ 0.014& CS\\
52530.0 &    ~~5.8 & $-$ & $-$0.020 $\pm$ 0.005 &  ~~0.002 $\pm$ 0.002 &  ~~0.046 $\pm$ 0.014& CS\\
52530.9 &    ~~6.7 & $-$0.005 $\pm$ 0.006 & $-$0.005 $\pm$ 0.006 & $-$0.001 $\pm$ 0.001 &  ~~0.034 $\pm$ 0.012& JJ\\
52531.5 &    ~~7.4 & $-$ & $-$0.006 & $-$0.001 &  ~~0.040& BA\\
52531.9 &    ~~7.8 & $-$0.004 $\pm$ 0.006 & $-$0.006 $\pm$ 0.006 & $-$0.001 $\pm$ 0.001 &  ~~0.041 $\pm$ 0.012& JJ\\
52533.5 &    ~~9.3 & $-$0.002 & $-$0.008 & $-$0.001 &  ~~0.042& BA\\
52533.9 &    ~~9.8 & $-$0.002 $\pm$ 0.006 & $-$0.009 $\pm$ 0.006 & $-$0.001 $\pm$ 0.001 &  ~~0.042 $\pm$ 0.012& JJ\\
52535.0 &   ~~10.9 &  ~~0.001 $\pm$ 0.005 & $-$0.029 $\pm$ 0.004 &  ~~0.000 $\pm$ 0.002 & $-$0.064 $\pm$ 0.010& DF\\
52537.9 &   ~~13.7 &  ~~0.001 $\pm$ 0.006 & $-$ & $-$ & $-$& JJ\\
52538.5 &   ~~14.3 &  ~~0.001 & $-$0.013 &  ~~0.000 &  ~~0.048& BA\\
52538.8 &   ~~14.7 & $-$ & $-$ & $-$ &  ~~0.048 $\pm$ 0.012& JJ\\
52539.9 &   ~~15.7 &  ~~0.001 $\pm$ 0.006 & $-$ & $-$ &  ~~0.049 $\pm$ 0.012& JJ\\
52540.0 &   ~~15.8 &  ~~0.004 $\pm$ 0.005 & $-$0.028 $\pm$ 0.004 &  ~~0.000 $\pm$ 0.002 & $-$0.050 $\pm$ 0.010& DF\\
52540.9 &   ~~16.7 &  ~~0.002 $\pm$ 0.006 & $-$ & $-$ &  ~~0.051 $\pm$ 0.012& JJ\\
52542.0 &   ~~17.8 &  ~~0.004 $\pm$ 0.005 & $-$0.025 $\pm$ 0.004 & $-$ & $-$& DF\\
52544.0 &   ~~19.9 &  ~~0.004 $\pm$ 0.005 & $-$0.022 $\pm$ 0.004 &  ~~0.000 $\pm$ 0.002 & $-$0.039 $\pm$ 0.010& DF\\
52544.9 &   ~~20.7 & $-$ & $-$0.014 $\pm$ 0.006 & $-$ & $-$& JJ\\
52545.9 &   ~~21.7 & $-$ & $-$0.014 $\pm$ 0.006 & $-$ & $-$& JJ\\
52546.5 &   ~~22.3 & $-$0.018 & $-$0.038 & $-$0.008 &  ~~0.024& BA\\
52546.8 &   ~~22.6 & $-$ & $-$ &  ~~0.002 &  ~~0.024& WM\\
52546.9 &   ~~22.7 & $-$ & $-$0.013 $\pm$ 0.006 & $-$ & $-$& JJ\\
52547.8 &   ~~23.6 & $-$ & $-$ &  ~~0.002 &  ~~0.042& WM\\
52547.8 &   ~~23.6 & $-$0.020 $\pm$ 0.006 & $-$0.037 $\pm$ 0.005 & $-$0.009 $\pm$ 0.002 &  ~~0.020 $\pm$ 0.012& AF\\
52548.9 &   ~~24.7 & $-$ & $-$0.011 $\pm$ 0.006 &  ~~0.002 $\pm$ 0.001 &  ~~0.037 $\pm$ 0.012& JJ\\
52551.5 &   ~~27.3 & $-$0.001 & $-$0.006 &  ~~0.002 &  $-$0.007& BA\\
52552.5 &   ~~28.3 & $-$0.011 & ~~0.000 & $-$0.024 & $-$0.017 & BA\\
52554.8 &   ~~30.7 & $-$0.013 $\pm$ 0.005 &  ~~0.008 $\pm$ 0.005 & $-$0.026 $\pm$ 0.006 &  ~~0.006 $\pm$ 0.010& MO\\
52555.8 &   ~~31.6 & $-$ & $-$ & $-$0.024 &  ~~0.016& WM\\
52555.8 &   ~~31.6 & $-$0.008 $\pm$ 0.006 &  ~~0.002 $\pm$ 0.006 & $-$0.024 $\pm$ 0.003 &  ~~0.016 $\pm$ 0.012& DO\\
52556.8 &   ~~32.6 & $-$ & $-$ &  ~~0.000 &  ~~0.016& WM\\
52560.4 &   ~~36.2 & $-$ & $-$0.002 $\pm$ 0.004 & $-$0.010 $\pm$ 0.003 & $-$0.050 $\pm$ 0.006& SS\\
52561.4 &   ~~37.2 & $-$0.012 $\pm$ 0.003 & $-$0.002 $\pm$ 0.004 & $-$ & $-$& SS\\
\hline
\multicolumn{7}{l}{($^a$) Relative to $B$ maximum (MJD= 52524.2)}\\
 \end{tabular}
\end{minipage}

\end{table*}

\section{Interstellar Extinction}

\begin{figure}
\includegraphics[angle=270, width=84mm]{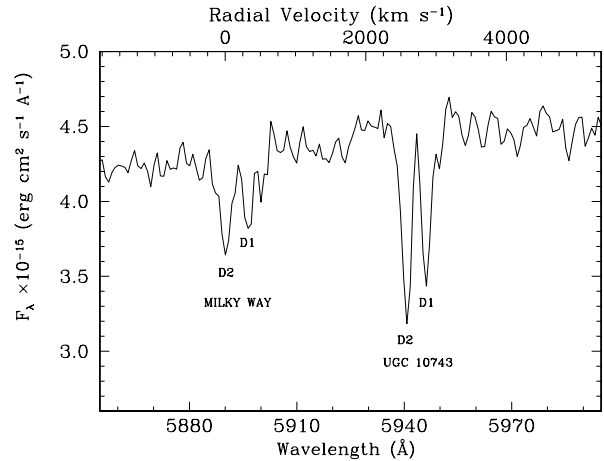}
\caption{Resolved NaI D1 and D2 lines relative to Milky Way and UGC~10743 present in the  medium-resolution spectrum of SN~2002er obtained on 2003, 
September 2 with the 2.5m Isaac Newton Telescope.}
\label{figure3}
\end{figure}

The presence of interstellar extinction toward SN~2002er is not
unexpected, since it appears projected onto the disk of UGC~10743 (see
Fig.~1).  Moreover, its relatively low galactic latitude
($b$=+26$^\circ$) suggests the presence of absorbing material in our
Galaxy too. Indeed, in the medium$-$resolution spectrum of SN~2002er
reported in Fig.~\ref{figure3}, the NaI D absorption lines are
evident  for both the Milky Way and the host galaxy.

In order to estimate the intrinsic luminosity of the SN it is crucial
to correct for extinction. In addition to the ``classical'' methods
reported in \citet{Phillips} and more recently proposed by
\citet{Wang} and \citet{Altavilla}, we have also estimated the colour
excess of SN~2002er  by comparing its colour evolution with that
of SN~1996X. These two events, in fact, were very similar to each other
as shown by the light curves (see Fig. \ref{figure4})
and spectra (Kotak et al. in preparation).

\begin{table}
 \caption{Reddening from different methods.}
 \begin{tabular}{@{}cccc}
  \hline
  Method & $E(B-V)$ & Reference \\
  \hline
$B_{max}-V_{max}$ & $0.34 \pm 0.05$ & \citet{Phillips}  \\
$V_{max}-I_{max}$ & $0.30 \pm 0.04$ & \citet{Phillips}  \\
$(B-V)_{max}$     & $0.33 \pm 0.06$ & \citet{Altavilla} \\
CMAGIC            & $0.31 \pm 0.04$ & \citet{Wang}      \\
$(B-V)_{tail}$    & $0.37 \pm 0.05$ & \citet{Phillips}  \\
$E(B-V)_{1996X}$  & $0.37 \pm 0.03$ & \citet{Salvo}     \\
  \hline
 \end{tabular}
\label{table5}
\end{table}

As pointed out by \citet{Leibundgut}, the derived absorption toward a SN
varies with the light-curve phase, due to the significant colour
evolution that these events undergo. Therefore the presence of the interstellar material
along the line of sight acts to change the shape of the light curve.
To take this effect into account, we have applied the correction around maximum brightness and in the
tail phase following the prescription given by \citet{Phillips}. As
in the comparison with SN~1996X, to account for the colour excess
evolution we need to compute a time dependent correction to be applied
to the observed $B-V$ of SN~2002er. For this purpose we have
artificially reddened a sample of archival SN spectra, originally
affected by a negligible amount of extinction, and  derived the
$E(B-V)$ evolution as a function of the SN phase.  The full results of
this work will be reported in a forthcoming paper (Pignata et al. in
preparation).

 Thanks to this procedure and to the S-correction that
we applied to the data, all the reddening estimates are consistent
within  $1\sigma$ (see Table~\ref{table5}).  The first four methods
listed in Table~\ref{table5} use maximum brightness data and only the
last two exploit  information carried by the tail and the whole
phase range, respectively.  Therefore, we have first computed the
weighted average of the $E(B-V)$ maximum values, $E(B-V)_{max}$, this
representing  the maximum light methods. Then
the final estimate of $E(B-V)$ was calculated as a weighted average of
$E(B-V)_{max}$, $E(B-V)_{tail}$ and $E(B-V)_{1996X}$. This gives $E(B-V) =0.36 \pm 0.05$.\\ 
\citet{Krisciunas2}  have presented $V-H$ and $V-K$ templates for ``Midrange'' decliners and they have proposed to use them as a tool to estimate the extinction suffered by a SN.
Using our single epoch IR observations (see section 5.3) we have obtained $E(B-V)=0.29 \pm 0.11$ and $E(B-V)=0.31 \pm 0.13$,  from the $V-H$ and $V-K$ colours, respectively. Both values are in reasonable agreement with those reported before. We have not included these estimates in the computation of  the $E(B-V)$ final value, because a single epoch measurement might introduce a systematic error if the observations were performed under non photometric conditions. Nevertheless, it tends to confirm that the reddening we have got from the  optical photometry is correct.

A totally independent way of estimating the extinction is given by the
relation between $E(B-V)$ and the equivalent width of the NaI~D lines,
as first proposed by \citet{Barbon}. More recently, \citet{Turatto1}
have shown that two different linear relations with different slopes
seem to exist. In the case of SN~2002er the relation with smaller
slope gives $E(B-V)=0.14$ for the Milky Way and $E(B-V)=0.42$
for the total amount of reddening. These values are in  good agreement
with the value reported in \citet{Schlegel} for the Galaxy alone (0.16) and 
the total value we get from the photometry (0.36), respectively.

\section{ Light and colour Curves}

\subsection{Maximum light}

\begin{figure*}
\includegraphics[width=150mm]{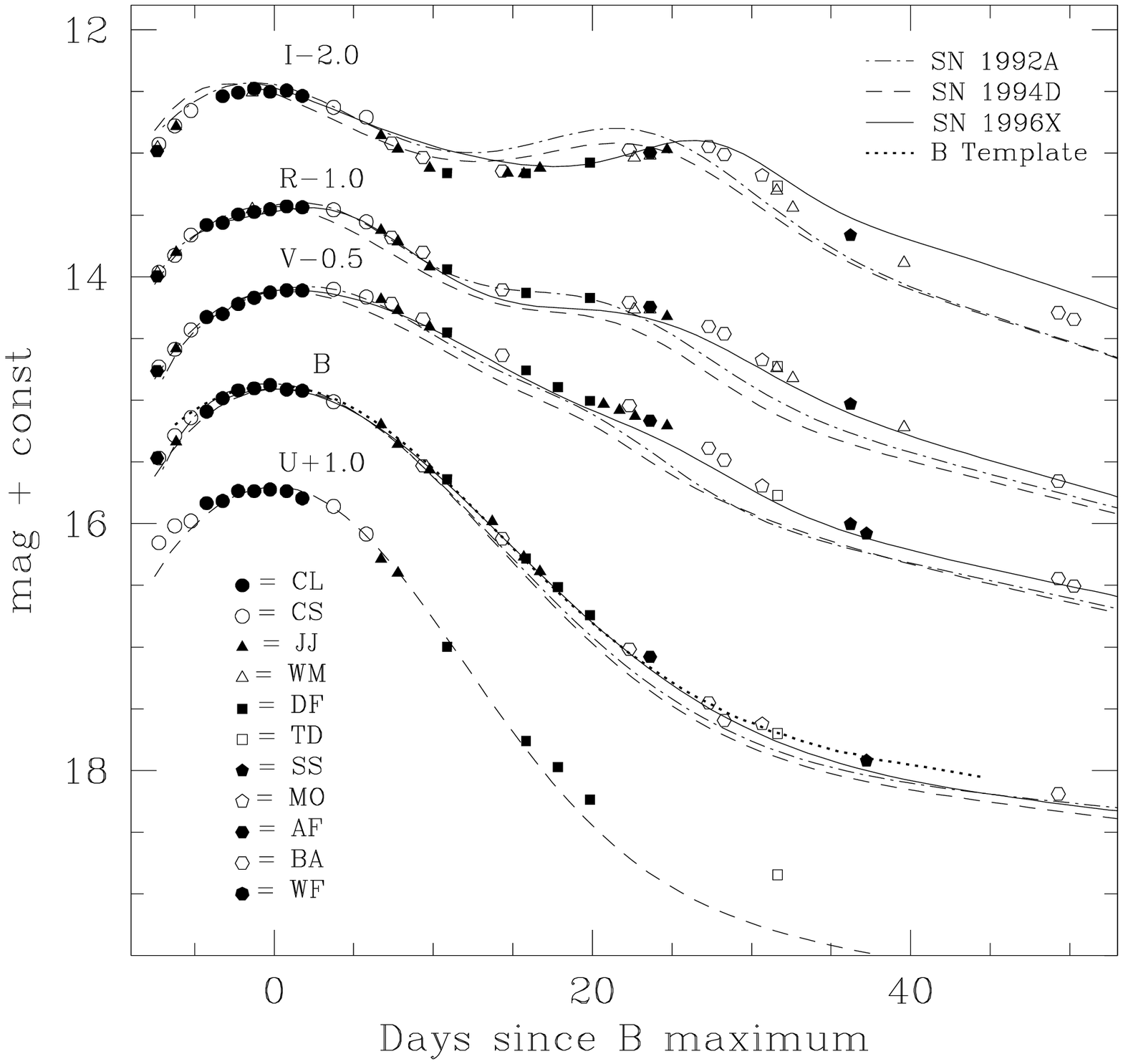}
\caption{$UBVRI$ S-corrected light curves of SN~2002er. The ordinate scale refers to 
the $B$-band. For presentation the other bands were shifted by the
amount shown in the plot. Different symbols refer to different
instruments (see Table \ref{table1} for the meaning of the
acronyms). The solid lines and dashed-dotted lines represent the
$BVRI$ light curves of SN~1996X ($\Delta m_{15}=1.31$) \citep{Salvo}
and SN~1992A ($\Delta m_{15}=1.47$), \citep{Suntzeff}. The dashed
lines refer to the $UBVRI$ light curves of 1994D ($\Delta
m_{15}=1.32$), \citep{Patat}. Finally the dotted line is the $B$ template
\citep{Leibundgut} stretched by a best fit factor $s=0.89$. The light curves of SN~1992A, SN~1994D, SN~1996X and the B template are shifted in order to get the best match to SN~2002er, between $-$7.0 and +7.0 days. }
\label{figure4}
\end{figure*}

The $UBVRI$ light curves of SN 2002er are shown in Fig.~\ref{figure4}.
For comparison, the light curves of three other Type Ia SNe with
similar values of $\Delta m_{15}$ are also sketched. From the plot it
is evident that SN~2002er resembles reasonably well all the three
template SNe, the match being the best with SN~1996X.  SN~2002er
follows the behaviour of this object even in the $I$ band, where the
differences between Type Ia SNe are usually more pronounced \citep{Suntzeff}.  

Our well sampled $UBVRI$ data give us the possibility of making a
detailed comparison with the other three SNe.  In the $U$ band the
light curve of SN~2002er shows a different behaviour with
respect to SN~1994D. In particular our object resembles
SN~1994D between $-$3 and +15 days, while later on  SN~2002er becomes
brighter. Unfortunately there are not a lot of observations of
SN~1996X in the $U$ filter, but the few available points are very
close to those of 1994D.  The $B$ light curve shows less pronounced
differences. For this band the SN Ia template \citep{Leibundgut}
stretched by a suitable factor
\citep{Perlmutter1} is also reported. The template fits the data reasonably
well over the whole phase range.  For the $V$ and $R$ filters we
notice that the shoulder visible between +20 and +30 days is
more pronounced, and starts slightly later than in SN~1994D and
SN~1992A, it is also stronger than SN~1996X.  In the $I$
light curve, the secondary maximum is achieved roughly at the same
epoch of SN~1996X and it is 0.53 mag fainter than the first maximum, a
value which is very similar to that of SN~1996X (0.52 mag).

As far as the pre-maximum phase is concerned, the comparison between SN~2002er
and SN~1994D seems to show a systematic shift in the rise time differences as we go from
the blue to the red passbands. The two SNe exhibit similar
behaviours in $V$ and $R$ filters, but SN~2002er reaches
maximum light more slowly than SN~1994D in the $U$ and $B$ bands and faster in the $I$ filter. 

The early discovery of SN~2002er also gives us the opportunity to apply
the method of \citep{Riess1} for estimation of the explosion
date ($t_0$). The main hypothesis of this method is that Type Ia SN shortly after the explosions could be represented as a expanding fireball. The SN  luminosity is therefore proportional  to the square time since the explosion. Combining the two unfiltered measurements
reported in the discovery IAU Circular (Li et al. 2002), which can be
reasonably well approximated to a $R$ band magnitude (Li, private
communication), with our early $R$ photometry, we obtain
$t_0(R)=52505.5$ (MJD). In this estimate we have considered only
points up to 8 days before the R maximum light, since, as shown by Riess and
collaborators, the reduced $\chi^2$ rises dramatically if later data
are included in the fit. Taking into account the 1.5 days delay
between $R$ and $B$ maxima, with the 7 available measurements we finally derive a rise time
$t_r(B)=18.7$ days, which is larger  than the $t_r=16.8$ days
implied by the relation of \citet{Riess1} given the $\Delta m_{15}$ of
SN~2002er.

  As mentioned before, reddening acts to
change the shape of the light curve, especially, in the $B$ filter, decreasing its decline
rate. To compute the reddening free $B$ band $\Delta m_{15}$, we
applied the correction found by \citet{Phillips} and obtained
$\Delta m_{15}= 1.33 \pm 0.04$.  Also the stretch factor was computed
correcting the $B$ light curve for the reddening effect using the time
dependent correction mentioned in the previous section. We note that
the reddening corrected values of $\Delta m_{15}$ and stretch factor
$s$, satisfy reasonably well both the relations reported by
\citet{Perlmutter2} and \citet{Altavilla}. 

The main photometric parameters of SN~2002er are reported in
Table~\ref{table6}.  Maximum light epochs and magnitudes were
estimated in each band fitting a low order polynomial. We note that
the time offsets between maxima in different passbands fall in the
range reported in \citet{Contardo}. In fact, SN~2002er reached
maximum brightness in the $U$ and $I$ band slightly earlier than in
the $B$ filter, but roughly two days later in $V$ and
$R$.

\begin{figure*}
\includegraphics[width=150mm]{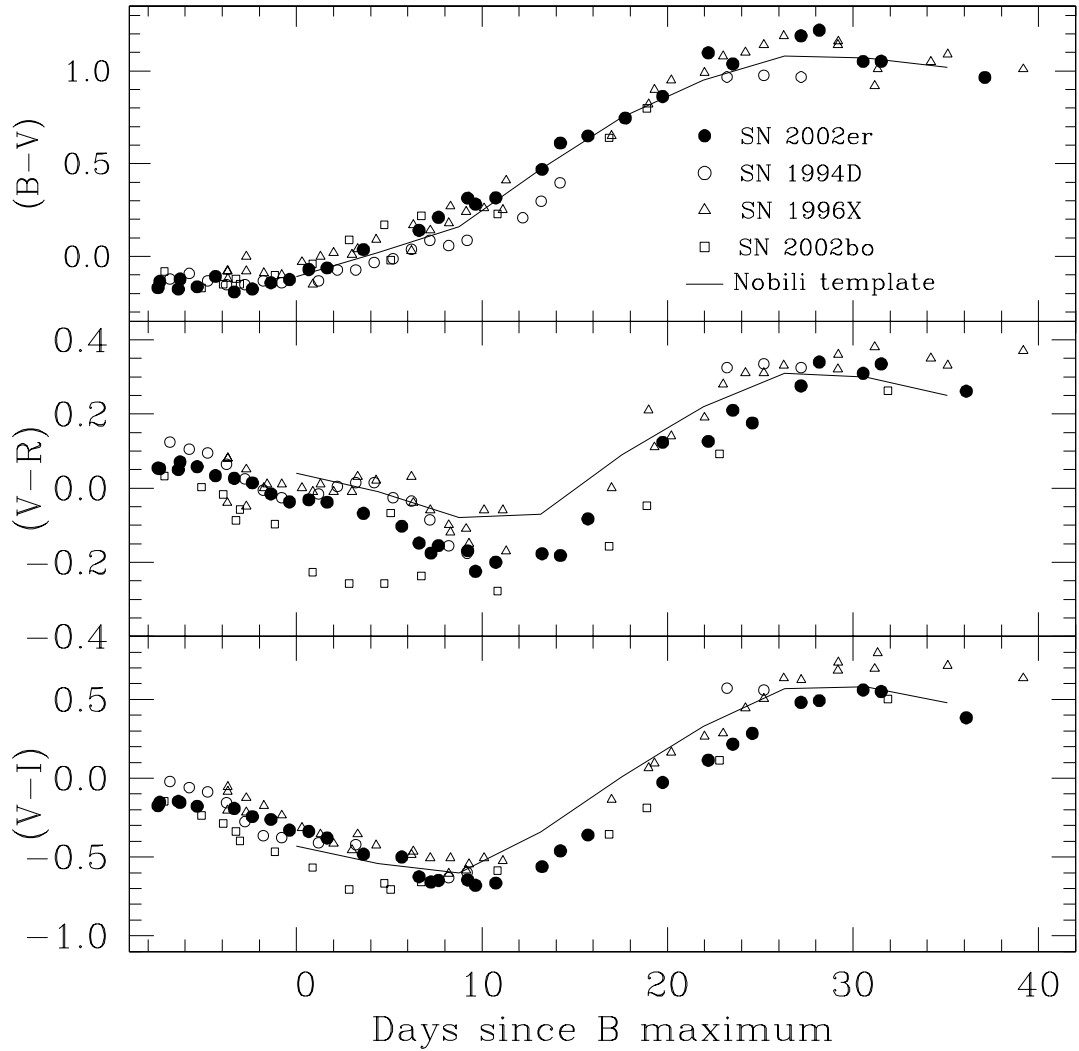}
\caption{$(B-V)_0$, $(V-R)_0$ and $(V-I)_0$ S-corrected colour curves of  SN~2002er 
compared with those of SN~1994D \citep{Patat}, SN~1996X \citep{Salvo},
SN 2002bo \citep{Benetti} and the Nobili colour curve template
\citep{Nobili} stretched by a factor $s$=0.89.}
\label{figure5}
\end{figure*}

\begin{figure*}
\includegraphics[width=150mm]{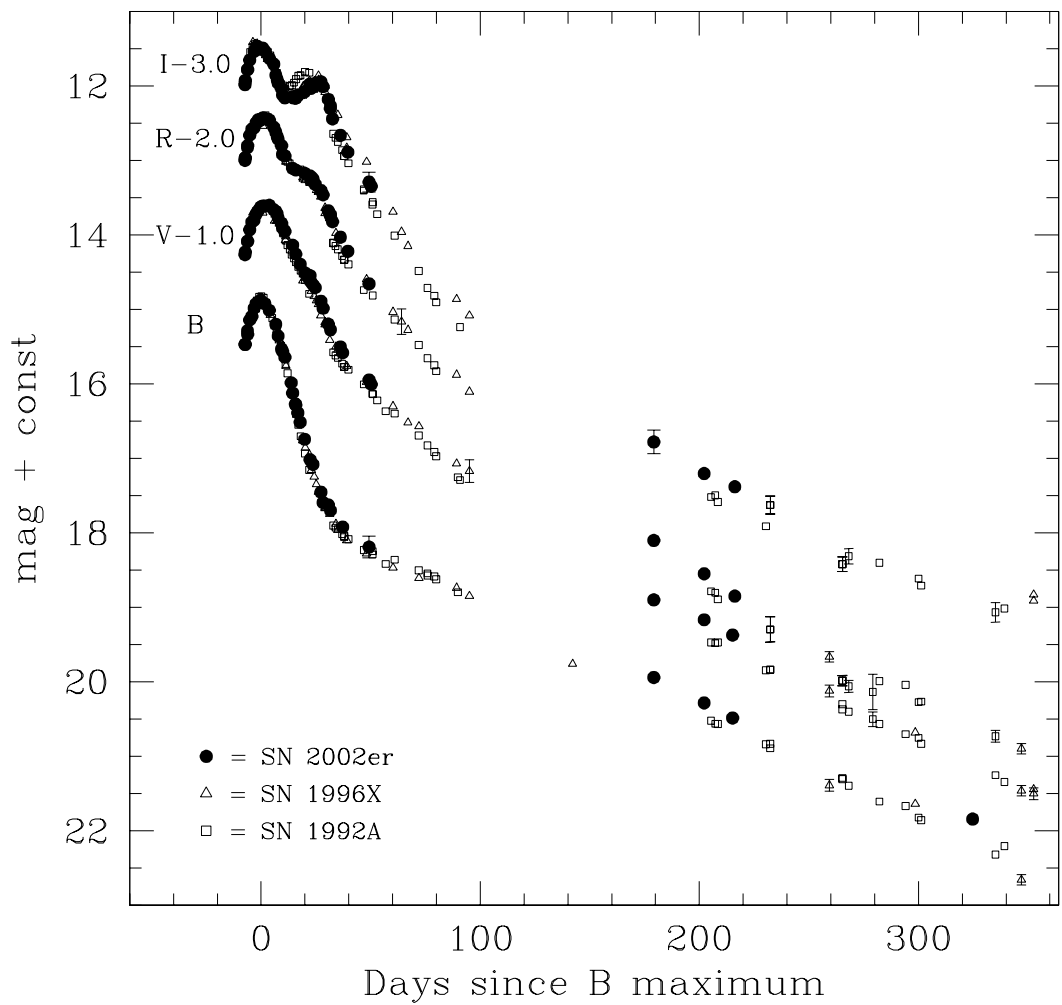}
\caption{$UBVRI$ light curves of SN~2002er (S-corrected) including late phase data (not S-corrected). The error bars are drawn only when they are larger than the points. 
For comparison, the light curves of SN~1996X \citep{Salvo} and
SN~1992A \citep{Suntzeff} are also sketched. The lack of data between
3 and 6 months is due to the seasonal gap. }
\label{figure6}
\end{figure*}

\begin{table}
 \caption{Main parameters of SN~2002er and its host galaxy. The
 decline rate $\gamma$ was computed taking into account the data later
 than +150 days only. All the values are computed using the S-corrected data.}
\begin{tabular}{@{}ll}
\hline
Host galaxy & UGC~10743 \\
Galaxy type & Sa? \citep{De Vaucouleurs} \\
 & Sc \citep{Christensen} \\
RA (2000) & 17$^h$11$^m$30$^s$.7\\
Dec (2000) &+07$^{\circ}$59'44''.8 \\
Recession velocity [km s$^{-1}$] & 2568~$\pm$~7 \citep{Falco}\\
& 2560~$\pm$~20 \citep{Christensen}\\
Recession velocity&  2652~$\pm$~33  \\
corrected for LG  infall onto & \\
 Virgo [km s$^{-1}$] & \\
Distance modulus & $\mu$~=~32.9~$\pm$~0.2 \\
($H_0$=71 km~s$^{-1}$~Mpc$^{-1}$) & \\
$E(B-V)$ & $0.36 \pm 0.05$  \\
Date of $B$ max (MJD) & 52524.2~$\pm$~0.5\\ 
Offset from the nucleus & $12''.3$  West and $4''.7$ North \\
Magnitude at max and & $U$~=~14.72 $\pm$ 0.04 ~~$-$0.5 [days] \\
time respect the $B$ max & $B$~=~14.89 $\pm$ 0.03 ~~~~0.0 [days]\\
& $V$~=~14.59 $\pm$ 0.03 ~~~~2.0 [days]\\
& $R$~=~14.43 $\pm$ 0.03 ~~~~1.5 [days] \\
& $I$~=~14.49 $\pm$ 0.05 ~~$-$0.6 [days]\\
Magnitude and epoch & ~~14.97 $\pm$ 0.05 ~~~~25.5 [days]\\
of the second $I$ max & \\
Reddening Corrected & $(U-B)_0=-0.42\pm0.08$ \\
Colours at time of $B$ max &  $(B-V)_0=-0.11\pm0.07$ \\
&  $(V-R)_0=-0.04\pm0.07$ \\
& $(V-I)_0=-0.32 \pm0.08$ \\ 
$\Delta m_{15}$ in $B$ & 1.33 $\pm$ 0.04\\
stretch factor in $B$ & 0.89 $\pm$ 0.02 \\
Late phase decline rate & $\gamma_B$=1.29~$\pm$~0.04 $\gamma_V$=1.48~$\pm$~0.04 \\
mag 100d$^{-1}$  &  $\gamma_R$=2.03~$\pm$~0.05 $\gamma_I$=1.67~$\pm$~0.11 \\
\hline 
\end{tabular}
\label{table6}
\end{table}

The de-reddened colour curves of SN~2002er are compared in
Fig.~\ref{figure5} with those of SN~1994D, SN~1996X, SN~2002bo and
with the colour template of \citet{Nobili} stretched by a factor
$s=0.89$. The  $(B-V)_0$ color evolution of SN~2002er is very similar to  all the other objects and it is well reproduced by the Nobili curve. In $(V-R)_0$ the differences between SNe are more
pronounced. In particular  between 0 and +8
days SN~2002er  shows neither the redder bump of the two events nor the bluer
dip of SN~2002bo. Indeed, in this phase interval, the $(V-R)_0$ evolution
is intermediate between the two cases. Anyway the overall curve
shape is quite well reproduced by the Nobili template, but
until +25 days it is about 0.1 mag bluer.

The $(V-I)_0$ colour evolution all the SNe return to be quite similar. We just note that until +9 days the SN~2002er $V-I$ evolution is very close to
 those of SN~1996X, SN~1994D and the Nobili template, while SN~2002bo
appears to be bluer. After +10 days, SN~2002er becomes bluer, reaching
almost the same colour as SN~2002bo.

\subsection{Nebular phase}

SN~2002er was observed also at late phases, i.e.  between six months
and one year after the explosion. The complete light curves to the latest epoch
are shown in Fig.~\ref{figure6}, together with those of SN~1992A and
SN~1996X. For presentation, the light curves of the last two objects
have been shifted to fit SN~2002er around maximum.  The $BVRI$ late time
decline rates $\gamma$, computed after day 150, are compatible within the errors
with those of SN~1992A and SN~1996X.  As found by \citet {Boisseau} at late epochs
the background contamination can play a relevant role and this could
be the reason why in each filter the points of SN~2002er lie about
0.1-0.2 magnitudes above those of SN~1992A and SN~1996X. This could be also the reason why in the last B band point this gap seems to became larger.
Another possible source of systematic errors might be related to the non-standard nature of the WFI filters. In fact, in this phase range the
emission features dominating the SN spectrum indeed act to amplify the
 effect of the passband mismatch. In particular, in Type Ia,  the strongest feature around 4700 \AA~identified as FeIII completely dominates the flux in the $B$ band. This, together with the fact that the WFI $B$ band departs considerably from the standard one, makes the $B$ S-correction for this instrument rather large. Using reddened  late spectra of SN~1996X and SN~1992A we found a value around 0.3 mag for SN~2002er. 
Also within the $V$ band there is quite a prominent feature at 5300 \AA~attributed to FeIII~+~FeII. However, in this case the $V$ WFI band is not so different from the Bessell one, so the   effect is quite small (0.05 magnitudes).
In the wavelength range covered by the $R$ and $I$ filters the spectrum is smoother than in the blue part. Therefore, even if the mismatch between the $R$, $I$ WFI bands and the $R$, $I$ Bessell bands is noticeable, the corrections are  below 0.1 magnitudes. 
Unfortunately  due to the low signal-to-noise ratio, the flux
calibration of these late time spectra is not precise enough to
compute a reliable correction.  However, since the spectral evolution
at late phases is very slow, the bands mismatch should not affect the
decline rate (see Table~\ref{table6}) which at this epoch  is the
most important photometric parameter.  
As mentioned by \citet{Turatto2}, the $B-V$ colour of Type Ia SNe
remains constant after six months from the peak brightness. SN~2002er
conforms to this behaviour: between day 179 and day 215 $B-V$ is
constant within the errors. The weighted mean of the three available
measurements is $B-V=0.09 \pm 0.04$.
At the present date the latest $B$ and $V$ imaging of SN~2002er was performed 619.1 days after maximum light. The goal of these observations was to detect a possible light echo. In fact, the detection of this rare phenomenon, observed only in two Type Ia SNe (SN~1991T, Schmidt et al. 1994 and SN~1998bu, Cappellaro et al. 2001), would allow one to get some information on the SN environment.
Using artificial stars placed at the SN position, we estimate a 3$\sigma$ upper limit of $B$=23.1 and $V$=23.7 for SN~2002er at the previously mentioned epoch.
Given these detection limits, we would have detected, for instance, a light echo similar to that shown by SN~1998bu, at least in the $V$ passband.
Of course we can not exclude a fainter echo, for example, as the one shown by SN~1991T.

\subsection{Near-IR Photometry}

A single epoch (19.9 days past $B$ maximum light) Near-IR imaging $(JHKs)$ was carried out at NTT Telescope using the SOFI IR camera on September 26.9 UT 2002.
This instrument uses a Hawaii HgCdTe 1024x1024 array and was operated at a plate scale of 0.29 arcsec/pixel. Data reduction was performed using standard routines in
IRAF. As in the Optical, IR  photometry was carried out  using the PSF fitting technique.
The calibrated magnitudes are $J=16.03\pm0.06$, $H=14.68\pm0.06$ and $Ks=14.66\pm0.07$. The associated errors are probably underestimate, since the observations were performed under unknown transparency conditions
Nevertheless the $(J-H)_0$ and $(H-K)_0$ colours are compatible within the uncertainties, with those of the \citet{Elias} IR templates.

\section{Absolute luminosity and bolometric light curves}

UGC~10743, the parent galaxy of SN~2002er, has a radial velocity with respect to the CMB reference frame of $v_r$=2573 $\pm$ 8 km s$^{-1}$ (LEDA).
To compute a reliable distance using the Hubble law, the recession velocity has to be corrected for motion departures from the Hubble flow.
One of the components of the peculiar motion is due to the Local Group (LG) infall into the Virgo Cluster. 
Using an infall velocity of 224 $\pm$ 90 km s$^{-1}$ \citep{Bureau} we obtain a corrected velocity of $v_r$=2652 $\pm$ 33 km s$^{-1}$ for UGC~10743.
Assuming $H_0$~=~71~$\pm$~8~km~s$^{-1}$~Mpc$^{-1}$ \citep{Freedman}, we derive
a distance modulus $\mu$~=~32.9~$\pm$~0.2. Taking into account the
estimated $E(B-V)$ and using the \citet{Cardelli} law of extinction in
the $UBVRI$ passbands, we obtain $M^{max}_{U}$~=~$-$19.9~$\pm$~0.3,
$M^{max}_{B}$~=~$-$19.5~$\pm$~0.2,
$M^{max}_{V}$~=~$-$19.4~$\pm$~0.2,
$M^{max}_{R}$~=~$-$19.3~$\pm$~0.2 and
$M^{max}_{I}$~=~$-$19.0~$\pm$~0.2.  Another way to estimate the
absolute magnitude is via the linear relations between $M^{max}$ and
$\Delta m_{15}$ first proposed by \citet{Phillips0}. This was later
revised, especially for the $B$ band, by other authors, the most
recent of whom is \citet{Altavilla}.  Adopting the \citet{Altavilla}
values -19.613~$\pm$~0.037 and 1.102~$\pm$~0.147 for the linear
coefficients, we obtain $M^{max}_{B}$~=~$-$19.35~$\pm$~0.07 for
SN~2002er, which is in quite good agreement with our previous estimate
of $M^{max}_{B}$.

\begin{figure}
\includegraphics[width=84mm]{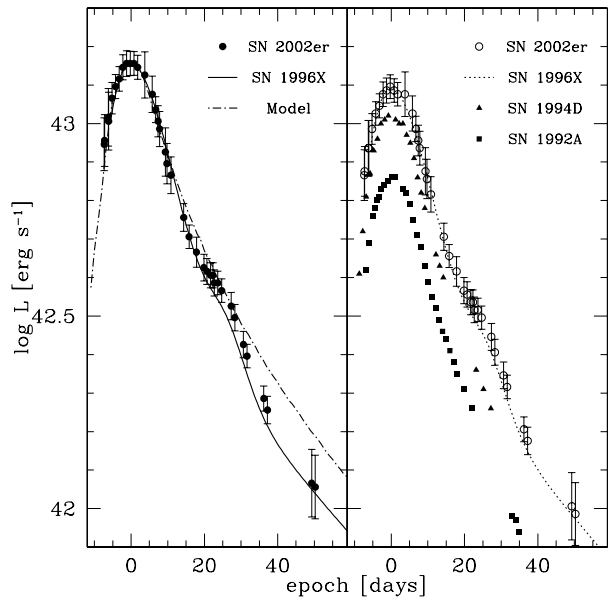}
\caption{ In the left panel the filled circles give the
{\it uvoir} light curve of SN~2002er computed using $R_{B}$ from
Cardelli et al. (1989) and $H_0$~=~71~$\pm$~8~km~s$^{-1}$~Mpc$^{-1}$ \citep{Freedman}. The
solid line is the {\it uvoir} light curve of SN~1996X shifted by +0.07
dex and the dashed-dotted line is a bolometric light curve model.  In
the right panel the open circles give the {\it uvoir} light curve of
SN~2002er computed using $R_{B}=3.5$ and
$H_0$~=~72~$\pm$~7~km~s$^{-1}$~Mpc$^{-1}$ \citep{Altavilla}. The
dashed line is the {\it uvoir} light curve of SN~1996X {\it without}
any shift relative to SN~2002er. The filled triangles and squares give
the {\it uvoir} light curves of SN~1994D and SN~1994A respectively.
Error bars refer only to the photometric errors and not to the
uncertainty in the reddening and distance.}
\label{figure7}
\end{figure}

Using the computed distance modulus and reddening and adding the UV
and IR contributions given by \citet{Suntzeff} to our well-sampled
$UBVRI$ data for SN~2002er, we were able to construct the $uvoir$
light curve. This is presented in Fig.~\ref{figure7}.  In the
left panel, we compare the $uvoir$ light curve of
SN~2002er (filled circles) with that of SN~1996X (solid line)
\citep{Riess_Salvo} shifted by +0.07 dex to obtain a best fit around
maximum.  The similarity in the $uvoir$ light curve shape between the
two objects is remarkable. SN~2002er shows only a longer rise time and a small deviation around the second maximum due to its more pronounced shoulder in the
red passbands. The 0.07 dex difference could be due to: (a) the uncertainties in the
distance modulus caused  by an error either on the Hubble flow recession velocity or on $H_0$ value (or a combination of both); (b) the
uncertainty in the reddening estimate; (c) an intrinsic difference in the mass of
$^{56}$Ni produced (or a combination of all three).  Concerning the
extinction, even if the colour excess is known with high accuracy, one
has to assume a value for the total-to-selective absorption
coefficient.  Typically, the values of Cardelli et al. (1989) are
adopted. Recently, \citet{Phillips}, \citet{Wang} and \cite{Altavilla}
have shown that for SN host galaxies a lower value of $R_{B}$
decreases the dispersion in the $M^{max}$ vs. $\Delta m_{15}$
relation.  In particular \cite{Altavilla}, adopting $R_{B}=3.5$ and
the Period-Luminosity-Colour relation of \citet{Freedman}, have
obtained $H_0$~=~72~$\pm$~7~km~s$^{-1}$~Mpc$^{-1}$.  These values of
$R_{B}$ and $H_0$ change the estimate of both the absorption and the
distance modulus of SN~2002er.  In the right panel of
Fig.~\ref{figure7} we show the $uvoir$ light curve (open circles) of
SN~2002er obtained assuming $H_0$~=~72~$\pm$~7~km~s$^{-1}$~Mpc$^{-1}$,
and adopting the total-to-selective absorption coefficients in the
$U$, $V$, $R$, and $I$ bands that correspond to $R_{B}=3.5$.  As the
reader can see, this curve resembles closely that of SN~1996X (dashed line) 
plotted without any shift relative to SN~2002er.  While this
might be seen as proof that the difference between the curves of SN~2002er and SN~1996X is due to reddening and distance uncertainties, the $uvoir$ light curves of other similar SNe are less supportive of this point of view.  Also
shown in the right panel of Fig.~\ref{figure7} are the $uvoir$ light
curves of SN~1992A, SN~1994D. Together with SN~1996X, these SNe have
well-determined distances, low extinction and similar $\Delta m_{15}$
values. Yet, in spite of these similarities, their luminosities are
quite different. In other words, in spite of the high similarity of
SN~1996X and 2002er, the $\Delta m_{15}$ value alone may not specify
the SN luminosity.

We have modelled the bolometric light curve of SN~2002er using our
Monte Carlo light curve code \citep{Mazzali}. The code follows the
emission, propagation, and deposition of the $\gamma$-rays and the
positrons emitted in the decay chain $^{56}$Ni$\rightarrow$
$^{56}$Co$\rightarrow$$^{56}$Fe, using constant $\gamma$-ray and
positron opacities $\kappa_{\gamma}=0.027$cm$^2$g$^{-1}$ and
$\kappa_{e^+}=7 $cm$^2$g$^{-1}$, respectively. It then follows the
random walk of the optical photons thus generated, adopting a time
dependent optical opacity appropriate for SNe~Ia as in
\citet{Mazzali}.  We used a W7 \citep{Nomoto} density/velocity structure to
characterise the SN ejecta.  In Fig.~\ref{figure7} (left
panel) we compare the model light curve ($^{56}$Ni mass set to
0.75$M_{\odot}$) with the observed $uvoir$ light curve of SN~2002er.
This model provides a good fit to the brightness and width near the
peak but fails to reproduce the bump seen at 25 days. This is probably
due to sudden changes in the opacity which our model does not take
into account at advanced stages.

The mass of $^{56}$Ni is comparable to the value estimated for SNe~Ia
having significantly slower decline rates. This suggests, as mentioned
before, that the estimated distance and/or reddening to SN~2002er may
have been overestimated. Alternatively (or in addition) it may be that
a single parameter is not enough to describe the type Ia event,
i.e. the $^{56}$Ni mass may not be as strongly correlated with the
light curve shape as has been suggested in the past.

\section{Light Curve Models}

Observed light curves and spectra provide the most direct test of
explosion models.
Here we present some synthetic light curves, not {\it fitted} to SN 2002er, to help interpreting
the data. For the interpretation of the colour light curves of SN~2002er (as well
as other Type Ia SNe) we will rely on the standard assumption
that they are white dwarfs, composed of carbon and oxygen, near the
Chandrasekhar mass, disrupted by thermonuclear burning. In the
``classical'' {\sl W7} model the velocity of
the thermonuclear burning front is parametrized in order to fit
the spectra of observed SNe Ia. More recent 3-dimensional   
explosion models (Reinecke, Hillebrandt \&. Niemeyer 2002a,b) are essentially free of
such non-physical parameters. They can only vary the ignition conditions
and the condition and composition of the exploding white dwarfs allowing more
self-consistent prediction of the light curves and spectra. Here we present synthetic
light curves based on both types of model, and compare them with
SN~2002er.

We use the light curve code STELLA  described in
\citet{Sorokina1}. It computes broad-band $UBVI$ (and
bolometric) light curves by solving (implicitly) the time-dependent
equations for the angular moments of the intensity in fixed frequency
bins, coupled to (Lagrangian) hydro-dynamics. In doing so no specific
temperature has to be ascribed to the radiation. The photon energy
distribution may be quite arbitrary.

While in this prescription the radiation is not in equilibrium
we assume LTE for ionization and level population. NLTE effects   
are simulated by using the approximation of the absorptive
opacity in spectral lines. In general, the effect of line
opacity is treated as an expansion opacity (Eastman \& Pinto 1993,
Sorokina \& Blinnikov 2002). It is obvious that these approximations
gradually loose reliability at late epochs when the SN enters
the nebular phase, and also are less accurate for the infrared
passbands in which the ejecta are more transparent. Therefore, here
we will concentrate mainly on the early $UBV$ light curves. 

\begin{figure*}
\includegraphics[width=150mm]{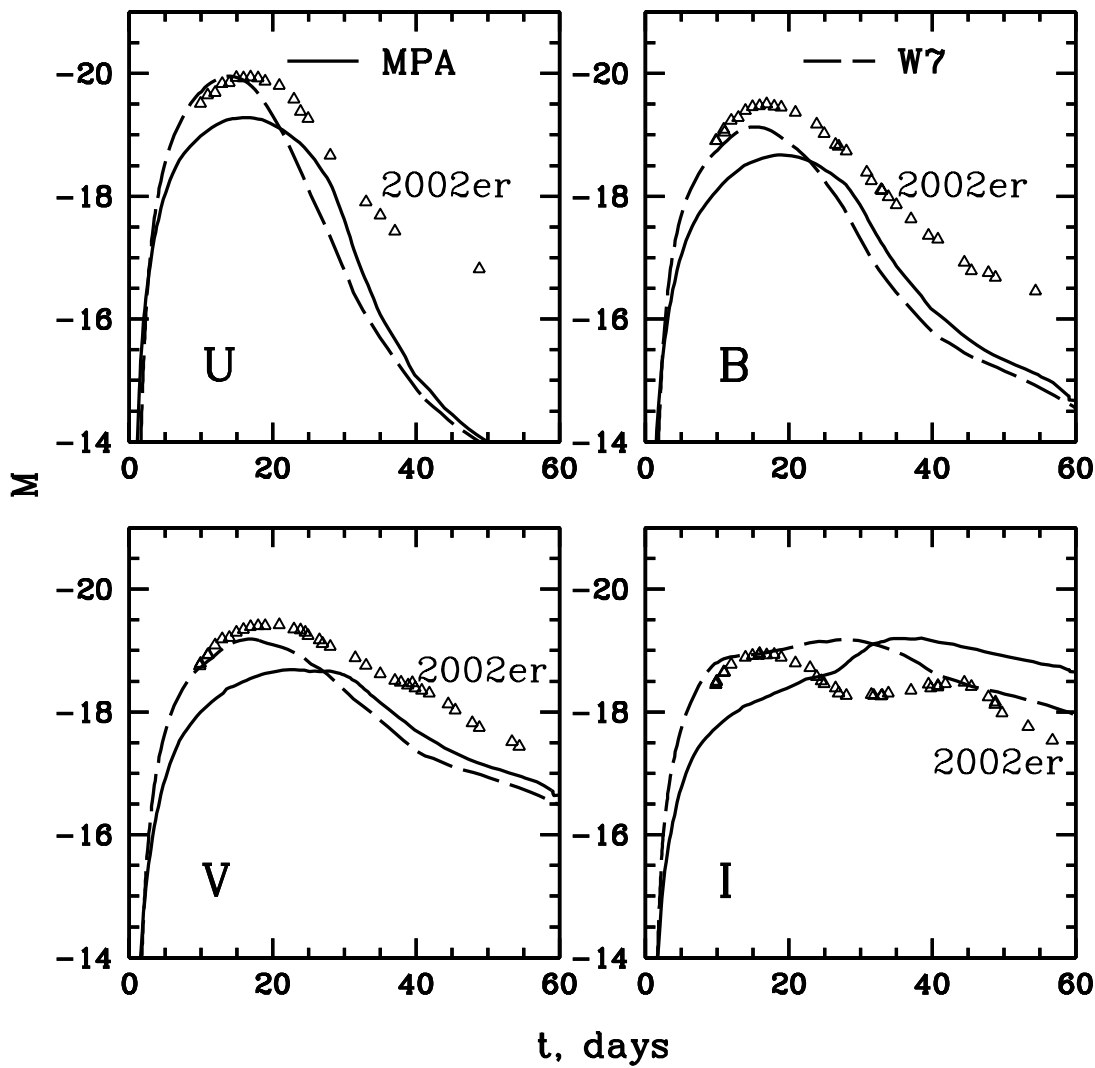}
\caption{UBVI-colour light curves predicted by a centrally ignited 3D
model (solid lines) and the spherically symmetric deflagration model
W7, compared with observed light curves of SN 2002er. Note
that no attempt was made to fit the observed data.}
\label{figure8}
\end{figure*}

Our results are shown in Fig.~\ref{figure8}. First of all
one can see that the W7-model reproduces rather nicely the rise of
SN~2002er in the $U$, $B$ and $V$ passbands, and also the peak luminosities
are in fair agreement with this model, which had a $^{56}$Ni-mass
of about 0.7 M$_{\odot}$. In contrast, the 3D model of \citet{Reineckeb}
with less $^{56}$Ni ($\simeq$ 0.4 M$_{\odot}$)
is too faint, but is in better agreement with the shape of the
light curves in $U$, $B$, and $V$. The rather poor fits to the I-band
light curve may reflect shortcomings of the light curve modelling,
as was discussed before.

The main reason for the broader light curves of the 3D supernova model,
as compared to W7, is the presence of high-velocity radioactive Ni in
the outer layers of the model which is not  predicted by the
spherical model W7.  Together with the fact that SN~2002er also showed
high-velocity Si and Ca in its early spectra (Kotak et al. in preparation) this
adds weight to the argument that the ejecta of SN~2002er, 
like those of SN 2002bo \citep{Benetti}, are heavily mixed.

\section{ Conclusions}

One of the main aims of the ESC project is the accurate and detailed
study of nearby Type Ia SNe in order to improve our knowledge of this
class of objects especially with respect to their use for cosmological
purposes.

The first step along this path is the construction of a data-base of
well-sampled light curves and spectroscopic data sets. SN~2002er is
one of the first targets of this observational campaign. For this
object, which was discovered about 14 days before $B$ maximum light,
we secured good spectroscopic coverage starting with day $-$11 and
this will be presented and discussed in a forthcoming paper (Kotak et
al., in preparation).  Here we have reported the results of the
photometric observations, which were carried out using a large set of
telescopes and instruments.

The $UBVRI$ measurements, which started on day $-$7.3, provided very
good sampling of the maximum light era and extended up to ten months
after the explosion. The ESC project aims at a high level of
observational precision and we have made particular efforts to correct
for the departures of the instrumental photometric systems from the
Johnson-Cousins standard system. This analysis has shown that in some
cases this can cause time- and SN spectrum-dependent systematic
deviations which can indeed hinder the cross-comparison with other
SNe.  In this respect, we note that the $I$ passband shows the largest
number of variations on the Cousins theme.  We note also that
published Type~Ia SN light curves show the largest variety in this
passband \citep{Suntzeff}. Therefore, some of the diversity seen
in this filter might be due to instrumental effects, a suspicion which
certainly needs more investigation, including cross-checks with the
spectral appearance of single objects in this wavelength range.

One of the problems we had to solve with SN~2002er is the extinction
correction. Interstellar absorption was expected due to the location
of the SN on the disk of the edge-on host spiral galaxy. Indeed,
clear signs of interstellar absorption was present in the
classification spectrum in the form of NaI~D absorption lines
\citep{Smartt}. This was confirmed by the colour curves analysis which
led to $E(B-V)$~=~0.36~$\pm$~0.05.\\
Dereddening the light curves,
assuming a standard total-to-selective absorption coefficient
$R_{B}=4.1$, we obtain a peak bolometric luminosity for SN~2002er of
$\log L$~=~43.2~$\pm$~0.2 erg s$^{-1}$ which is 0.07 dex (i.e. $\sim$17\%) larger
than that of SN~1996X. This is in spite of the very similar shapes of
the bolometric light curves of the two SNe.  However, by setting
$R_{B}=3.5$ for the host galaxy, we obtain excellent agreement for both the light curve
shapes and luminosities.  The impressive resemblance to SN~1996X
suggests that the artificial shift we had to apply to match the two
curves is due to distance and/or reddening errors.  However, we do not
rule out the possibility that the luminosities of SN~2002er and
SN~1996X might be intrinsically different. This suspicion is
strengthened when we examine the bolometric light curves of SN~1992A,
SN~1994D and SN~1996X. These SNe are characterized by low extinction,
better constrained distances and similar $\Delta m_{15}$, and yet
their absolute luminosities are quite different.  We conclude that the
single parameter characterisation of the Type Ia SN may sometimes lead
to incorrect luminosity estimates i.e.  the $^{56}$Ni mass may not be
as strongly correlated with the light curve shape as has been
suggested.  More information about the similarity between SN~2002er
and SN~1996X will be provided by the spectroscopic analysis, which
through synthetic spectral modelling, will also give independent
distance and reddening estimates (Kotak et al., in preparation).

We have compared the $UBVI$ light curves of SN~2002er using two type of
explosion codes: $W7$ from Nomoto et al. (1984) and the 3-D models from
Reinecke et al. (2002b).  $W7$ seems to better reproduce the rise and
peak luminosity of the SN~2002er light curves, but it fails in the
post maximum phase. The 3-D explosion model appears to be
systematically fainter than the observed data, but better reproduces
the overall shape.  The 3-D model predicts 0.4 M$_{\odot}$ of
$^{56}$Ni, while $W7$ produces a larger amount (0.7 M$_{\odot}$). The
latter $^{56}$Ni-mass value is in good agreement with the value
obtained by modelling the bolometric light curve (0.75 M$_{\odot}$).
As mentioned in section~6, such a large $^{56}$Ni mass is more common
for slower-declining SNe~Ia. The $^{56}$Ni mass could be overestimated
due to errors in the distance and/or absorption values, but it may
also imply that the correlation with the light curve shape is not so
strong.

Although SN~2002er could be defined as a ``normal'' Type Ia SN, some
small differences, not affected by the uncertainties in distance and
reddening, have been highlighted thanks to the excellent photometric
coverage.  Comparison of the pre-maximum phase of SN~2002er with that
of SN~1994D shows that, despite the two SNe having similar $\Delta
m_{15}$, SN~2002er rises more slowly in the $U$ and $B$ bands but
faster in the $I$ filter.  The $B$ light curves of SN~2002er and
SN~1996X are almost indistinguishable until 30 days after the $B$
maximum light, but the shoulder between 15 and 30 days in the $V$ and
$R$ bands is clearly more pronounced in SN~2002er than in
SN~1996X. SN~2002er also exhibits a deeper and earlier minimum in the
I~filter than does SN~1996X.  The $(V-R)_0$ curves show an even
larger dispersion: on the one hand one has {\it normal} events like
SN~1994D or SN~1996X and on the other deviant objects like SN~2002bo
($\Delta m_{15}$~$\sim$~1.1), which at any time is clearly bluer
(see for example Fig.~6). SN~2002er shows an intermediate behaviour.

All these facts tend to support the emerging idea that a single
parameter description is not sufficient to fully characterise Type Ia
SNe (see the discussion in Benetti et al. 2004).

\section*{Acknowledgments}
We thank Enrico Capellaro for carefully reading the manuscript and for his advise on photometric data reduction. Stephen J. Smartt thanks PPARC for the financial assistance.
This work is partially based on observations made with ESO Telescopes
at the La Silla Observatory under programme ID 169.D-0670, ID 59.A-9004  and Paranal Observatory under programme ID 073.D-0853. It is
supported in part by the European Community's Human Potential
Programme under contract HPRN-CT-2002-00303, ``The Physics of Type Ia
Supernovae''. This work is also based on observations performed at Calar Alto
Observatory, Spain, the Jacobus Kapteyn Telescope (JKT) of the Isaac
Newton Group, La Palma, the Beijing Observatory, China, the
Wendelstein Observatory, Germany, by Otto Baernbantner and Christoph Ries,
 the Italian Telescopio Nazionale Galileo (TNG), La
Palma, the Asiago Observatory, Italy.  The TNG is operated on the
island of La Palma by the Centro Galileo Galilei of INAF (Istituto
Nazionale di Astrofisica) at the Spanish Observatorio del Roque de los
Muchachos of the Instituto de Astrofisica de Canarias.  This work has
made use of the NASA/IPAC Extragalatic database (NED) which is
operated by the JET Propulsion Laboratory, California Institute of
Technology, under contract with the National Aeronautic and Space
Administration. We have also made use of the Lyon-Meudon Extragalatic
Database (LEDA), supplied by the LEDA team at the Centre de Recherche
Astronomique de Lyon, Observatoire de Lyon.

 \bsp \label{lastpage} \end{document}